\documentclass[10pt,aps,twocolumn,showpacs,superscriptaddress,prl]{revtex4-1}
\usepackage{graphicx}
\usepackage{amsfonts,amssymb}
\usepackage{enumerate}
\usepackage[sort&compress]{natbib}
\usepackage{hyperref}
\usepackage[latin1]{inputenc}

\usepackage{amsfonts,amsmath,amssymb,graphicx,color,amsthm}
\usepackage{tgtermes}

\begin{document}

\title{Coexistence of ballistic and Fourier regimes in the $\beta$-FPUT lattice}

\author{Giovanni \surname{Dematteis}}
\affiliation{Dipartimento di Fisica, Universit\`a degli Studi di Torino, Via Pietro Giuria 1, 10125 Torino, Italy}

\author{Lamberto \surname{Rondoni}} 
\affiliation{Dipartimento di Scienze Matematiche, Politecnico di
  Torino, Corso Duca degli Abruzzi 24, I-10129 Torino, Italy}

\author{Davide  \surname{Proment}}
\affiliation{School of Mathematics, University of East Anglia, Norwich Research Park, NR47TJ Norwich, United Kingdom} 

\author{Francesco \surname{De Vita}}
\affiliation{Dipartimento di Fisica, Universit\`a degli Studi di Torino, Via Pietro Giuria 1, 10125 Torino, Italy} 

\author{Miguel \surname{Onorato}}
\affiliation{Dipartimento di Fisica, Universit\`a degli Studi di Torino, Via Pietro Giuria 1, 10125 Torino, Italy}
\affiliation{INFN, Sezione di Torino, Via Pietro Giuria 1, 10125 Torino, Italy}

\date{\today}

\begin{abstract}

Commonly, thermal transport properties of one-dimensional systems are found to be anomalous. 
Here, we perform a numerical and theoretical study of the $\beta$-FPUT chain, considered a 
prototypical model for  one-dimensional anharmonic crystals, in contact with thermostats at different temperatures. We give evidence that,  in steady state conditions, the {\it local} wave energy spectrum can be naturally split into modes that are essentially ballistic (non-interacting or scarcely interacting) and  kinetic modes (interacting enough to relax to local thermodynamic equilibrium).  We show numerically that the well-known divergence of the energy conductivity 
is related to how the transition region between these two sets of modes shifts in $k$-space with the system size $L$, due to properties of the collision integral of the system. 
Moreover, we show that the kinetic modes are responsible for a macroscopic behavior compatible with Fourier's law.
Our work sheds light on the long-standing problem of the applicability of standard thermodynamics in one-dimensional nonlinear chains,  testbed for understanding the thermal properties of nanotubes and nanowires.

\end{abstract}

%
%
%


\maketitle

Deriving equations for the macroscopic observables from microscopic dynamics is the core of statistical physics. Often, such ``reduction'' passes through an intermediate stochastic model for the mesoscopic scale, c.f. the celebrated Boltzmann equation for a rarefied gas~\cite{cercignani1972boltzmann}, used to derive the hydrodynamic equations (e.g. Euler, or Navier-Stokes)~\cite{de1989incompressible}. Linear response theory is a sound framework to compute the macroscopic transport coefficients~\cite{kubo1957statistical,kreuzer1981nonequilibrium,callen1998thermodynamics}.
In a solid rod whose ends are kept at not excessively different temperatures one expects Fourier's law of heat conduction to hold and the Green-Kubo formula to yield the conductivity as an integral of the time-correlation of the heat current at equilibrium~\cite{kubo1957statistical}.
This program has been pursued for one-dimensional (1D) systems~\cite{rieder1967properties,lepri2003thermal,dhar2008heat}, typically modelled as lattices of point particles interacting via spring-like forces.
In fact, 1D structures such as
nanowires and nanotubes have become available
in lab experiments, with widespread applications
in industry and technology~\cite{chang2008breakdown,chen2010remarkable,shen2010polyethylene,yang2010violation,liu2012anomalous}, making an understanding of 1D transport of primary importance. Thus, harmonic~\cite{rieder1967properties} as well as anharmonic lattices have been widely studied~\cite{dhar2008heat,lepri2016thermal}.
Nevertheless, establishing the macroscopic equations for low-dimensional structures results challenging~\cite{dhar2008heat,lepri2016thermal}; the dimensional constraints imply a slow decay of correlations that may prevent the convergence of the Green-Kubo integral.  The approach based on the wave kinetic equation, i.e. the phonon Boltzmann equation of solid state physics and main object of wave turbulence theory~\cite{nazarenko2011wave}, has recently opened an important perspective in this field~\cite{pereverzev2003fermi,spohn2006phonon,lukkarinen2016kinetic,onorato2015route}. The wave kinetic equation concerns phonons that interact with each other through resonant $n$-wave collisions, providing an effective relaxation mechanism toward thermodynamic equilibrium~\cite{onorato2015route,lvov2018double,pistone2018thermalization,pistone2018universal}. Although the heat conductivity can be computed rigorously ~\cite{aoki2006energy,spohn2006phonon,lukkarinen2016kinetic} when  a pinning onsite potential is introduced,
for unpinned anharmonic lattices energy conduction appears nontrivially anomalous~\cite{lepri1998anomalous,prosen2000momentum,pereverzev2003fermi,lukkarinen2008anomalous,dhar2008heat,lepri2016thermal}.
The usual procedure is to introduce a heuristic cut-off in the Green-Kubo integral based on the sound speed~\cite{lepri2016thermal,lukkarinen2016kinetic}, but a rigorous justification is still lacking~\cite{dhar2008heat}.
Moreover, there is no clear interpretation of the Green-Kubo formula in cases where the deviations from local thermodynamic equilibrium  are significant~\cite{aoki2003violations,lepri2005one,hurtado2006breakdown,giberti2011anomalies,giberti2017n,giberti2019temperature,mejia2019heat}. Mainly due to these difficulties, the derivation of proper macroscopic equations for such systems remains a long-standing open problem.

In this Letter, we use concepts of wave-kinetic theory to investigate the low-temperature regime of the $\beta$ Fermi-Pasta-Ulam-Tsingou model ($\beta$-FPUT )~\cite{fermi1955alamos,gallavotti2007fermi,dauxois2008fermi,lepri2005studies}, paradigmatic anharmonic 1D lattice. In the thermodynamic limit, the mechanism of thermalization at the mesoscopic scale is related to 4-wave resonant interactions \cite{pistone2018universal}.
We give  evidence from direct numerical simulations that the system splits into two independent sets of modes: {the low-$k$ modes, with mean-free-path exceeding what we call the ``mesoscopic'' scale $\lambda$, and the remaining modes that, interacting resonantly, relax to local thermodynamic equilibrium.
 We show that the high-$k$ modes transport energy diffusively, hence the anomalous-transport scaling is due to how the wave numbers associated with the transition modes (not diffusive but also not perfectly ballistic) scales with $L$. 
More precisely, we give evidence that (i) Our scaling of the transition between ballistic and kinetic modes dictates the scaling of the energy conductivity with the size $L$}; (ii) The conductivity restricted to the kinetic modes converges, as in standard heat conduction; (iii) The energy density is dominated by the locally thermalized kinetic modes, implying an apparently regular Fourier temperature profile. Results are supported by extensive numerical simulations and a dimensional argument applied on the wave kinetic equation.

The $\beta$-FPUT system is defined by the Hamiltonian $\mathcal{H} = \mathcal{H}_h + \mathcal{H}_a$ with
\begin{equation}\label{eq:0}
\mathcal{H}_h = \sum_{j=1}^N\frac12 p_j^2 +\frac12 (q_j-q_{j-1})^2, \;\; \mathcal{H}_a = \frac{\beta}{4}\sum_{j=1}^N (q_j-q_{j-1})^4.
\end{equation}
Let $J_e$ be the (harmonic) energy current per particle,
\begin{equation}\label{eq:0bis}
\begin{aligned}
 J_e = \frac{1}{N}\sum_{j=1}^N \frac12 (p_j+p_{j-1})(q_j-q_{j-1})
\end{aligned}
\end{equation}
$T_\pm=\bar T\pm\Delta T/2$ the temperatures of the thermostats
at the two ends of the $\beta$-FPUT
chain, where $N$ is the number of particles of unitary mass and unitary lattice spacing ($N=L$). Let us assume $\Delta T \ll \bar T$ and consider the low-temperature regime, so that the ratio of the anharmonic and the harmonic parts of the Hamiltonian is small, $\mathcal{H}_{a}/\mathcal{H}_h\simeq \beta \bar T\ll1$. For lattices, it is standard to define the thermal conductivity as
\begin{equation}\label{eq:1}
\kappa_e =  \langle J_e\rangle N/\Delta T\,,
\end{equation}
where $\langle \cdot \rangle$ indicates canonical ensemble average, assuming that the system is close to equilibrium at temperature $\bar T$. 
For normal heat conduction $\kappa_e$ tends to a constant as $N\to\infty$,
while transport is called anomalous if $\kappa_e$ diverges, e.g. 
as $N^\delta$, $\delta > 0$.
 For the harmonic chain ($\beta=0$) the problem was solved analytically in~\cite{rieder1967properties}:
the conductivity is proportional to $N$
($\delta = 1$).
Such transport behavior, with
no temperature gradient arising in the chain bulk (since wave packets, or {\it phonons}, do not interact), is called ballistic.
On the other hand, for the $\beta$-FPUT chain
 an exponent $\delta\simeq 0.4$ was found by some authors, while others proposing $\delta=1/2$ or $\delta = 1/3$~\cite{lepri2016thermal}.

In the wave-turbulence formalism ~\cite{nazarenko2011wave,pistone2018universal,aoki2006energy,spohn2006phonon} we use the normal variable $a_k :=(\omega_k q_k + i p_k )/\sqrt{2\omega_k}$,
with $\omega_k=2|\sin(k/2)|$, and the wave-action spectral density associated with $a_k$,
$n_k=n(k,x,t)$. One can derive, although not fully
rigorously, the following wave kinetic equation
~\cite{pereverzev2003fermi,spohn2006phonon,lukkarinen2008anomalous,nazarenko2011wave,lukkarinen2016kinetic,onorato2015route,lvov2018double}:
\begin{equation}\label{eq:3}
\frac{\partial n_k}{\partial t} + v_k\frac{\partial n_k}{\partial x}=I_k.
\end{equation}
The collision integral $I_k$ is conveniently split as a difference  of 
two terms:
$I_k=\eta_k - n_k/\tau_k$, with
\begin{equation}\label{eq:4}
\begin{aligned}
&\eta_{k_1}= {4\pi} \int_{-\pi}^{\pi} |T_{1234}|^2n_{k_2}n_{k_3}n_{k_4}\delta(k_{12}^{34})\delta(\omega_{12}^{34})dk_2dk_3dk_4\,, \\
&\frac{1}{\tau_{k_1}}={4\pi} \int_{-\pi}^{\pi} |T_{1234}|^2(-n_{k_3}n_{k_4}+n_{k_2}n_{k_3}+n_{k_2}n_{k_4})\\
&\qquad\qquad\qquad \times\delta(k_{12}^{34})\delta(\omega_{12}^{34})dk_2dk_3dk_4 \,,\\
& |T_{1234}|^2 = \frac9{16}\omega_{k_1}\omega_{k_2}\omega_{k_3}\omega_{k_4},
\end{aligned}
\end{equation}
in a spatial domain  $x\in[0,L]$, using the notation: $y_{12}^{34}:=y_{k_1}+y_{k_2}-y_{k_3}-y_{k_4}$.
The group velocity is denoted by $v_k=d\omega_k/dk$. The second term in the l.h.s. of equation~\eqref{eq:3} quantifies advection due to spatial inhomogeneities. 
$\tau_k$ can be interpreted as the mean collision time of
mode $k$.

In the harmonic chain, the collision integral is missing and  Eq.~\eqref{eq:3}  predicts ballistic advection of wave-action carried by harmonic excitations with speed $v_k$. The r.h.s. of~\eqref{eq:3}
is the 4-wave collision integral, and represents the effective mechanism of relaxation to equilibrium~\cite{pistone2018universal}. Note that, for finite $N$ and small nonlinearity, 6-wave resonant interactions dominate \cite{onorato2015route,lvov2018double,bustamante2019exact}, but we made sure that $N$ in our simulations is large enough for the Fourier space to be sufficiently dense, thereby making 4-wave interactions dominate~\cite{pistone2018universal}.

Now, $n_k$ must be a slowly varying function of $x$:
Eq.~\eqref{eq:3} assumes the existence of a mesoscopic scale, $\lambda$, much smaller than the macroscopic scale. This suggests an operative definition of $\lambda$ as the largest size over which the system can be considered approximately spatially homogeneous, i.e. as the largest scale on which the use of the Fourier transform is justified. When the system is thermodynamic, the characteristic microscopic interaction distance is finite. Thus, a mesoscopic scale $\lambda$ within which
practically all phonons interact, and relaxation takes place, can be
identified~\cite{kreuzer1981nonequilibrium}. Moreover, boxes of size $\lambda$ appear as points of
a continuum in the $\lambda/L \to 0$ limit. 
In short, considering a partition of the system consisting of $L/\lambda$ adjacent boxes, two limits need to be taken: (i) $L \to \infty$ at fixed boundary temperatures, so that the object is large compared
to the microscopic scales and a continuum description makes sense; (ii) $\lambda\to\infty$, that yields the continuous $k$-space
formalism (thermodynamic limit) of~\eqref{eq:3}, at fixed $x$. In principle, there are different ways of combining these limits, that correspond to different physical situations.
In our theoretical argument,  we will assume that the separation  between the mesoscale and the macroscale is of one order of magnitude, i.e.  $L\propto\lambda^2$;
 such choice corresponds to the mesoscopic scale being placed  in between the microscopic and the macroscopic scales ({\it cf}. the hydrodynamic diffusive scaling of the Boltzmann equation~\cite{de1989incompressible,saint2009hydrodynamic}).
We will give evidence that our results, obtained integrating numerically the $\beta$-FPUT equation of motion, are consistent with our assumption.

Having defined the scales, we focus our attention  on the wave kinetic equation~(\ref{eq:3}). In stationary conditions, there are two competing contributions: the collision integral and the transport term. In order to observe regular heat conduction, we expect the collision term to guarantee local relaxation to thermodynamic equilibrium within any homogeneous subdomain of width $\lambda$. On the other hand, if the collision integral is not active enough, energy flows in a ballistic way from one subdomain to the next one, without having thermalized. As observed in our numerical simulations,
the situation for the $\beta$-FPUT system is hybrid: the collision integral is able to make the system thermalize locally up to some critical wave number $k_c$, while lower $k$ waves scarcely interact and carry energy almost
ballistically between adjacent subdomains. Using a dimensional argument, the scaling of $k_c$ with $L$ can be estimated from the wave kinetic equation assuming that, in stationary conditions, the transport term and the collision integral balance:
\begin{equation}
\frac{v_{k_c}}{\lambda}\sim \frac{1}{\tau_{k_c}}\,.
\label{eq:dimen}
\end{equation}
Assuming very small deviations from local equilibrium (see discussion at the end of the letter) 
and small $k$, a direct analytical calculation~\cite{pereverzev2003fermi,lukkarinen2008anomalous} yields
\begin{equation}\label{eq:100}
\tau_k\propto k^{-5/3}\,, \quad \text{for}\quad |k|\ll1\,,
\end{equation}
which plugged into eq. (\ref{eq:dimen}) leads to
\begin{equation}
k_c\propto\lambda^{-3/5}\propto L^{-3/10},
\label{eq:10}
\end{equation}
where we have used the fact that $L\propto \lambda^2$ and, for small $k$, $v_k \simeq $ 1.
Based on the above discussion, we conjecture that:  i) the field $a_k$ can be split into two parts: the modes with $|k| > k_c$ are essentially in thermodynamic equilibrium within cells of size $\lambda$ and verify Fourier's law, while for $|k|\le k_c$ the modes are ballistic or scarcely interacting;  ii) the scaling of $\kappa_e$ with $L$ is strictly related to the scaling of $k_c$ with $L$, Eq.~\eqref{eq:10}. Numerical evidence of the conjecture is given below.

%
\begin{figure}
\includegraphics[width=0.9\linewidth]{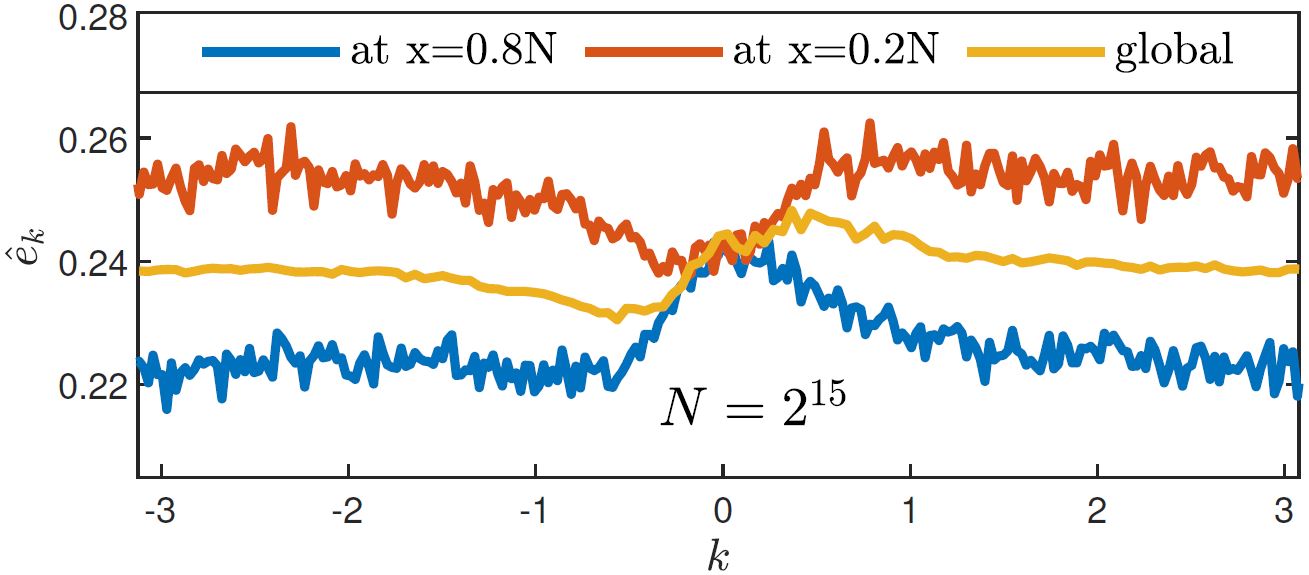}%
  \caption{\small Steady state energy spectrum for $N=2^{15}$.
   \label{fig:0}}
\end{figure}
\begin{figure}
\includegraphics[width=0.9\linewidth]{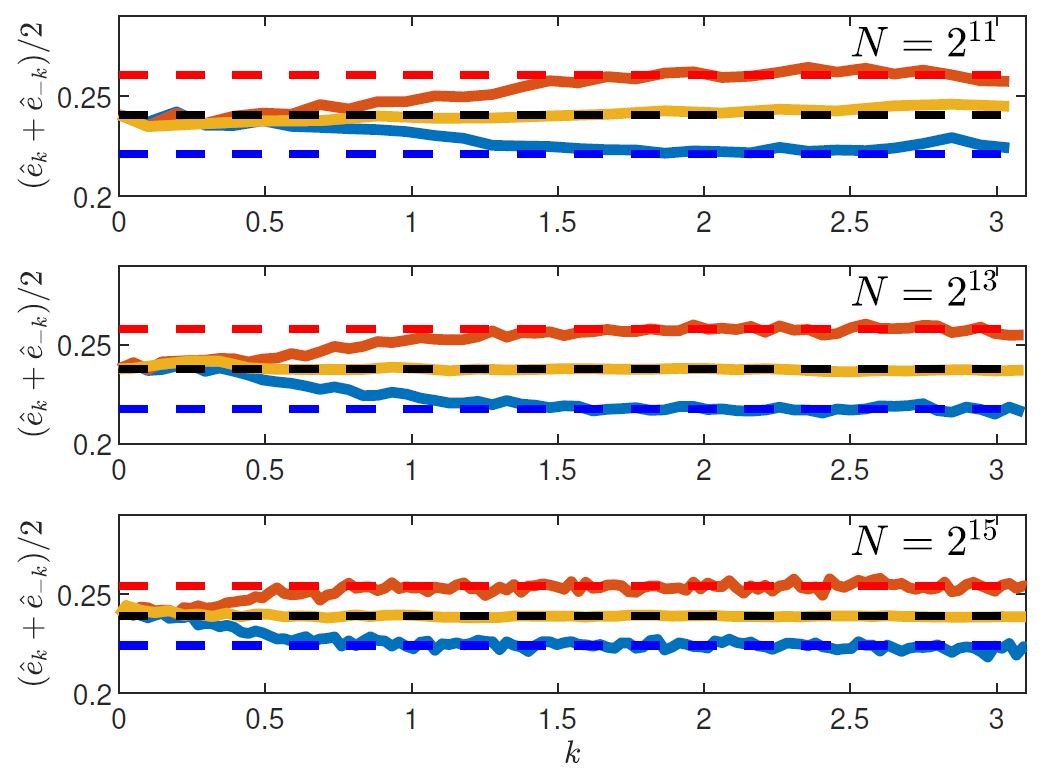}%
  \caption{\small The red and blue solid lines are the symmetrized stationary {local energy spectra computed}  in windows of width $\sqrt N$ centered at $x_1=0.2N$ and $x_2=0.8N$, respectively. The dashed lines with same colors are at the respective average energy per particle at $x_1$ and $x_2$. The yellow line is the {spectrum performed on the full chain}.
   \label{fig:1}}
\end{figure}
Our simulations solve $2N$ coupled equations of motion for the Hamiltonian~\eqref{eq:0} with $\beta=0.1$,
integrated with fourth order Runge-Kutta method with time step $5\times10^{-2}$. Fixed boundary conditions are used for the two walls $q_0=0$ and $q_{N+1}=0$, and the two particles $j=1$ and $j=N$ are coupled with Nos\'e-Hoover thermostats at temperatures $T_+=0.3$ and $T_-=0.2$, respectively at the left and the right ends. In all our simulations we consider $N$   sufficiently large so that the discrete-$k$ effects, which lead to a lack of four-wave resonances, do not take place, see  \cite{pistone2018universal}.  Differently from previous works where the spectral analysis is performed on the full length of the chain,  we are interested in the local spectral properties of the system. Therefore, we consider  the  mesoscopic length $\lambda=\sqrt{N}=\sqrt{L}$. Macroscopic observables are averaged spatially over mesoscopic boxes of size $\lambda$, and in time at the steady state, over $2.5\times10^5$ time units after an initial relaxation transient of $5\times10^4$ units. An additional ensemble averaging over $5$ independent realizations is used to improve statistical convergence.
In Fig.~\ref{fig:0}, we show the 
wave energy spectral density (energy spectrum), $\hat e_k=\omega_k n_k$, computed locally on a spatial
window of mesoscopic size,  around two points placed at
$x = 0.2L$, close to the thermostat at higher temperature and $x = 0.8L$, close 
to the one at lower temperature. Observe that the energy spectrum is clearly asymmetric for small $k$.
 Moreover, for  high wave numbers the spectrum is almost flat, i.e. those modes are in an equipartition state. In order to highlight local thermalization, noticing that the asymmetric part of the spectrum does not contribute to the total energy, in Fig.~\ref{fig:1} we plot the symmetrized energy spectrum $(\hat e_k + \hat e_{-k})/2$,  for different values of $N$. Local equipartition is observed for the high wave numbers, say the modes with $|k|>k_c$. Instead, the energetic content of the low-wave-number modes ($|k|<k_c$) tends to the average temperature $\bar T$ throughout the chain.
As $N$ increases, $k_c$ shifts toward the origin and the separation between
the two states becomes sharper. For $N = 2^{15}$, the two states are
clearly separated, with a narrow transition region.

\begin{figure}
\includegraphics[width=0.9\linewidth]{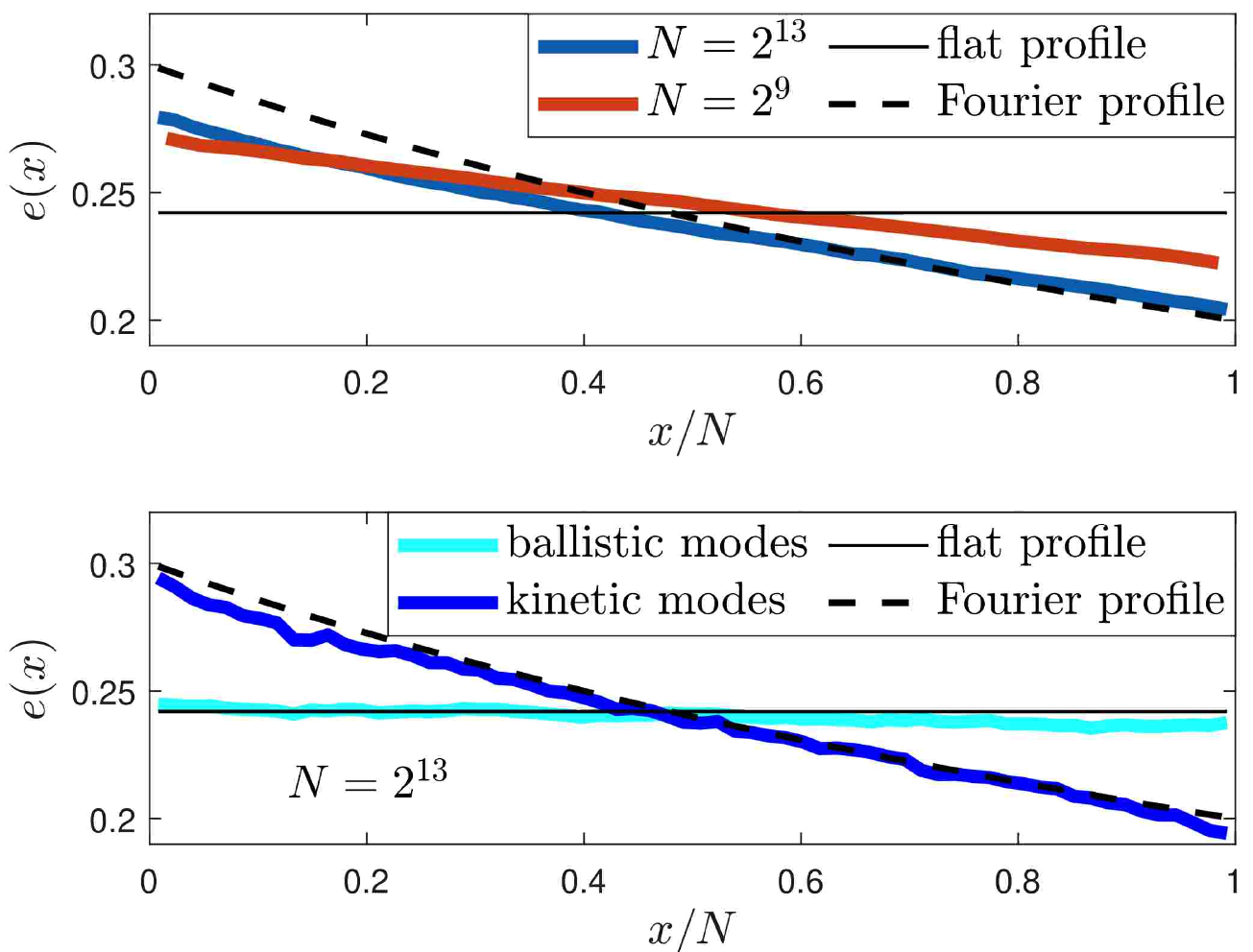}
  \caption{\small Top: Numerical profiles of $e (x)$. Bottom: The blue line of the upper panel ($N=2^{13}$) is decomposed into its contributions from $|k|\le k_c$ (light blue) and $|k|>k_c$ (dark blue), normalized by the fraction of modes in each set: $k_c/\pi$ and $(\pi-k_c)/\pi$, respectively. A value $k_c=0.75$ is used, based on numerical estimate from Fig.~\ref{fig:1}.
 \label{fig:2}}
\end{figure}

In Fig.~\ref{fig:2} we show  the energy per particle $e(x)$, averaged in time and in space over mesoscopic boxes of width $\lambda=\sqrt{N}$ for $N=2^9$ and $N=2^{13}$. The figure shows that as $N$ increases the profile tends to the expected Fourier profile. In the lower panel, the case for $N=2^{13}$ is further investigated and $e(x)$ is decomposed as
\begin{equation}
e(x)=e^>(x;k_c)+e^<(x;k_c)
\end{equation}
using the estimate $k_c=0.75$ from the central panel of Fig.~\ref{fig:1}, and
where $e^>(x;k')$ is obtained from  $q(x)$ and $p(x)$ after they have been filtered  in Fourier space  and only contributions from wave numbers $|k|> k'$ have been retained;  $e^<(x;k')$  has contributions only from $|k|\le k'$. As conjectured, the profile for $e^>(x;k_c)$ is consistent with the expected linear profile, typical of Fourier's law, and the second one is consistent with a flat profile with temperature $\bar T$; this latter behaviour is typical of harmonic (collisionless) chains. 


\begin{figure}
\includegraphics[width=\linewidth]{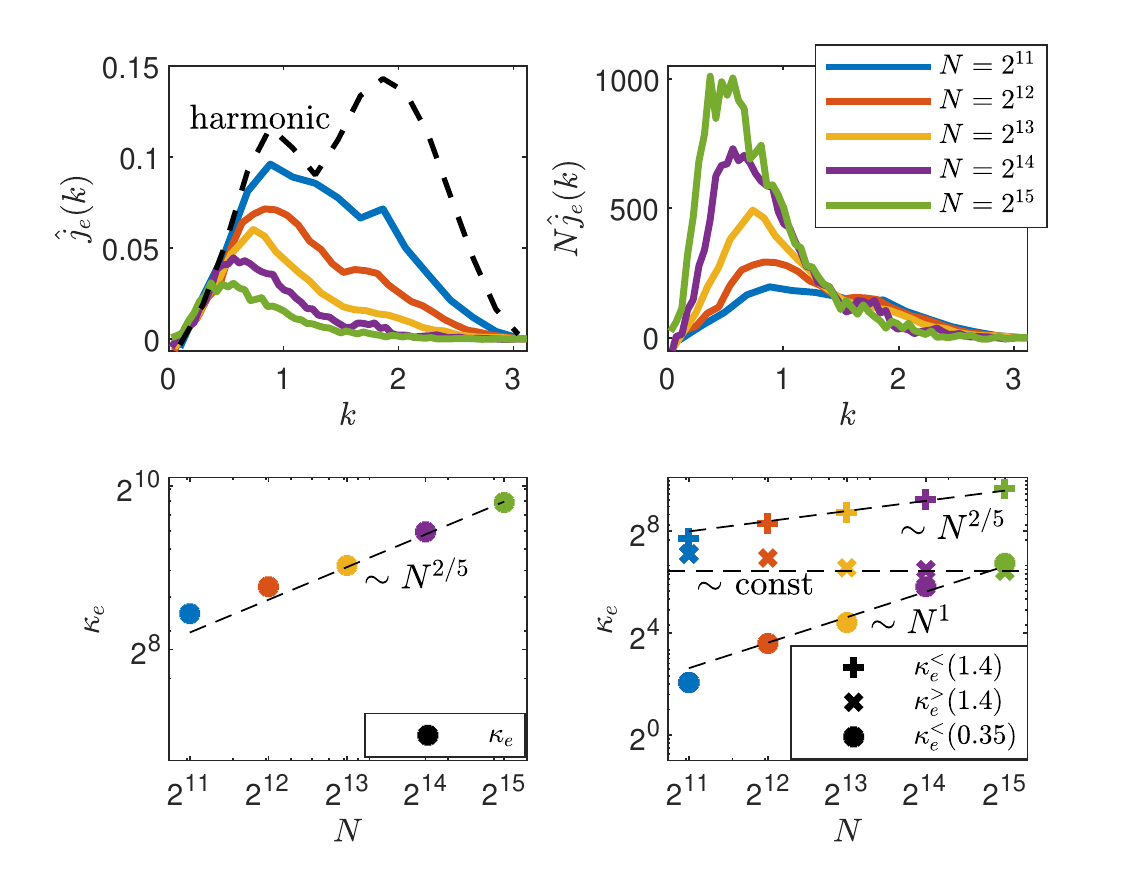}
  \caption{\small Top left: net average energy current per mode $\hat j_{e}(k)$. Top right: scaled current $N\hat j_{e}(k)$ (cf. Fig. 2.6 of~\cite{dhar2016heat}). Bottom left: $N$-dependent scaling of $\kappa_e$ compared to the $N^{2/5}$ scaling (black dashed line). Bottom right: $N$-dependent scaling behavior for the filtered conductivity: ballistic $O(N)$ divergence at low $k$'s, kinetic constant conductivity at high $k$'s, and $O(N^{2/5})$ divergence. 
 \label{fig:3}}
\end{figure}
We now devote our attention to the energy fluxes; in steady state conditions the flux is independent of $x$. We therefore define the net spectral energy current~\cite{lepri2003thermal},  as
\begin{equation}\label{eq:18}
\begin{split}
&\hat  j_{e}(k) := v_k\omega_k(n_{k}-n_{-k})/2\,,
\end{split}
\end{equation}
which depends on the asymmetric part of the spectrum. Note that $\hat j_{e}(k)=\hat j_{e}(-k)$ and the Fourier transform for calculating $n_{k}$ is made on the entire length of the chain. This is because in stationary conditions the flux is independent of $x$.
In Fig.~\ref{fig:3}, the upper panels show the behavior of $\hat j_e(k)$ while varying $N$.
As $N$ doubles, $ \hat j_{e}$ remains constant for low wave numbers (ballistic modes), while it roughly halves for higher wave numbers (kinetic modes). This means that  the ballistic-modes contribution to the energy current is independent of $N$, as expected for the harmonic chain~\cite{rieder1967properties}, while for the kinetic modes it is proportional to $N^{-1}$. In the lower left panel we show the energy conductivity as defined by Eq.~\eqref{eq:1}, where $\langle J_e\rangle=\sum_k \hat j_e(k) $. The plot shows a scaling compatible with previous numerical results, $\kappa_e\sim L^{2/5}$. We now consider the filtered conductivity
\begin{equation}\label{eq:18}
\kappa_e^{>}(k)=\tfrac{N}{\Delta T}\sum_{|k'|>k}\hat j_e(k'),\; \kappa_e^{<}(k)=\tfrac{N}{\Delta T}\sum_{|k'|<k}\hat j_e(k').
\end{equation}
with $\kappa_e =  \kappa_e^<(k) + \kappa_e^>(k)$.
On the lower right panel of Fig.~\ref{fig:3} we show three quantities with different scalings: Choosing a threshold $k=1.4>k_c$ for all $N$ (cf. Fig.~\ref{fig:1}), we observe that $\kappa_e^<(1.4)\sim N^{2/5}$, while $\kappa_e^>(1.4)$ converges to constant (compatible with regular Fourier's law). On the other hand, considering wave numbers that are always ballistic ($0.35<k_c$ for all $N$), thus excluding modes around $k_c$, we obtain that $\kappa_e^<(0.35)\sim N^{1}$, typical of the harmonic chain but far from the anomalous exponent $0.4$. Thus, the way $k_c$ scales with the size $L$, Eq.~\eqref{eq:10}, plays a key role concerning the anomaly, as will be shown in the following.
Multiplying~\eqref{eq:3} by $\omega_k$ and integrating over $k\in[-\pi,\pi]$ yields
\begin{equation}\label{eq:14}
{\partial_\tau} e(x,\tau) = -{\partial_x} j_e (x,\tau),
\end{equation}
where $e (x,\tau):=\int_{-\pi}^{+\pi}\hat e(k) dk \,$ and $j_e (x,\tau):=\int_{-\pi}^{+\pi}\hat j_e(k) dk \,$ are the macroscopic energy density and current, respectively.
 Let us split the current as
\begin{equation}\label{eq:15}
 j_e (x,\tau)= j_e^<(x,\tau;k_c) + j_e^>(x,\tau;k_c)\,;
\end{equation}
we then consider $ j_e^< = 2\int_0^{k_c}\hat j_e(k) dk \propto k_c^2 $ --- since $\hat j_e(k)\propto k$ for $k\ll1$, see  Fig.~\ref{fig:3} top left --- and $j_e^>=-\kappa_{\rm kin}\partial_x e$, with $k_{\rm kin}>0$ --- i.e. Fick's law for the kinetic modes. Using the estimate of Eq.~\eqref{eq:10}, $k_c\sim L^{-3/10}$, we obtain:
\begin{equation}\label{eq:16}
 \kappa_e^<(k_c)  \propto L^{2/5}\,, \kappa_e^>(k_c)=\kappa_{\rm kin}\,.
\end{equation}
Plugging~\eqref{eq:15} into~\eqref{eq:14} and taking into account that in the bulk the energy associated to ballistic modes is constant, we obtain
\begin{equation}\label{eq:17}
	\partial_{x}\left(\kappa_{\rm kin}\;\partial_{x} e^> (x;k_c)\right) = 0\,.
\end{equation}
Imposing the fraction $1-\tfrac{k_c}{\pi}$ of kinetic modes to have ``temperature'' $T_{\pm}$ at the boundaries; and that the fraction of ballistic modes $\tfrac{k_c}{\pi}$ is at constant ``temperature'' $\bar T$, we have
\begin{equation}\label{eq:19}
 e(x) = e^<(x;k_c) + e^>(x;k_c)= \tfrac{k_c}{\pi} \bar T + \left(1-\tfrac{k_c}{\pi} \right)T(x)\,,
\end{equation}
where $T(x) = T_+ -  \tfrac{\Delta T}{L}x$ is the Fourier profile~\cite{spohn2006phonon}, attained by the kinetic part as solution to~\eqref{eq:17}. 
Thus, Eqs.~\eqref{eq:16}-\eqref{eq:19} highlight how the kinetic modes dominate the energy density, despite the presence of an anomalous conduction.

{\it Discussion. }
Our numerical simulations  show that the $\beta$-FPUT lattice does not thermalize locally; moreover, the scaling obtained for 
 $\kappa_e$ is compatible with the law $L^{2/5}$.
We have developed a theoretical argument, based on dimensional analysis
of the kinetic equation, that supports this view. Because other
numerical simulations and theoretical approaches have led to different
exponents~\cite{lepri2016thermal},
we mention that the key ingredient in our approach is that the
relaxation time, calculated from the collision integral, scales as
$k^{-5/3}$ for small $k$. Such result is exact if computed at equilibrium,
i.e. for a thermalized spectrum, hence it is subject to not necessarily
universal corrections in nonequilibrium conditions. To verify
its robustness, we have performed a direct numerical computation of the collision integral and we have calculated the relaxation time scale in the presence of different kind of perturbations, including also the introduction of a chemical potential.
The results are contained in the Supplemental Material and can be summarized as follows: for small perturbations, the $k^{-5/3}$  scaling is confirmed; however, as the perturbation grows, a departure   from the $k^{-5/3}$ scaling, especially in the very low wavenumber region, is observed.

To conclude, strictly speaking, the $\beta$-FPUT lattice does not relax to local equilibrium, since only part of its normal modes reaches such state. The system can be treated as made of two independent components, with distinct macroscopic behavior: The high-$k$ modes satisfying Fourier's law, with linear temperature profile and regular heat transport with finite heat conductivity; The low-$k$ modes, scarcely interacting, carrying energy in a ballistic way (or close to ballistic): this differs in essence from heat conduction~\cite{chen2013diffusion}.
Our conclusions are in qualitative agreement with the efficient transport due to low frequency modes in carbon nanotubes, offering a simple first-principle interpretation to the persistence of coherent ballistically-conducting modes observed experimentally at lengths up to more than $0.1\,${\it mm}~\cite{donadio2016simulation,zhang2012low,chang2016experimental}, about $10^5$ times the tube diameter and $10^6$ times the interatomic distance.



\bibliographystyle{apsrev4-1}
\bibliographystyle{apsrev4-1}

\pagebreak
\section*{Supplemental Material to the Letter:  ``Coexistence of ballistic and Fourier regimes in the $\beta$-FPUT lattice''}

The homogeneous Wave Kinetic equation, also known as the phonon Boltzmann equation, associated to the $\beta$-FPUT equation, in  wave-action spectral density variables,
$n_k=n(k,x,t)$, has the following form \cite{nazarenko2011wave}:
\begin{equation}
\begin{aligned}
&\frac{\partial n_{k_1}}{\partial t} = {4\pi} \int_{-\pi}^{\pi} |T_{1234}|^2n_{k_1}n_{k_2}n_{k_3}n_{k_4}\\
&\left(
\frac{1}{n_{k_1}}+\frac{1}{n_{k_2}}-\frac{1}{n_{k_3}}-\frac{1}{n_{k_4}}
\right)
\delta(\Delta K)
\delta(\Delta \Omega)dk_2 dk_3 dk_4 
\label{eq:WK}
\end{aligned}
\end{equation}
where $\Delta K=k_1+k_2-k_3-k_4$ and  $\Delta \Omega=\omega(k_1)+\omega(k_2)-\omega(k_3)-\omega(k_4)$.
It is known that such kinetic equation has the following stationary solution:
\begin{equation}
n_{RJ}(k)=\frac{T}{\omega(k)+\mu},
\label{RJ}
\end{equation}
known as the Rayleigh-Jeans solution. The constant $T$, the temperature, and the chemical potential, $\mu$, are
associated with the conservation of linear energy and number of quasi-particles, respectively. The case $\mu=0$ corresponds to the equipartition of energy. Strictly speaking, neither the energy, nor the number of particles, are conserved quantities of the original microscopic equation of motion. Indeed, the $\beta$-FPUT conserves the full Hamiltonian (sum of a quadratic and a quartic part), while the Wave Kinetic equation only the energy associated with the quadratic part (in its statistical version). We note that the estimated value of $\mu$ in the simulations in this Letter is $\mu=O(0.01)$. This justifies energy equipartition ($\mu=0$) as a good approximation, but the role of $\mu$ in general and its relation with the properties of the thermostat will be the topic of further studies.

We have numerically solved the collision integral using the standard procedure of removing the two $\delta$-functions and, consequently, reducing the 3-dimensional integral into a 1-dimensional one. The procedure is based on the existence of a non trivial resonant manifold and on the determination of the loci that compose such manifold (theorems on the subject can be found in \cite{lukkarinen2016kinetic}). %
We have verified numerically that an arbitrary initial condition relaxes to the predicted Rayleigh-Jeans distribution (details will be published elsewhere). Of major interest for the present Letter is the calculation of the relaxation time scale $\tau_k$, defined as
\begin{equation}
\begin{aligned}
&\tau_{k_1}=\bigg({4\pi} \int_{-\pi}^{\pi} |T_{1234}|^2(-n_{k_3}n_{k_4}+n_{k_2}n_{k_3}+n_{k_2}n_{k_4})\\
&\qquad\qquad\delta(\Delta K)\delta(\Delta \Omega)dk_2dk_3dk_4\bigg)^{-1}
\end{aligned}
\end{equation}
\begin{figure}
\includegraphics[width=0.5\textwidth]{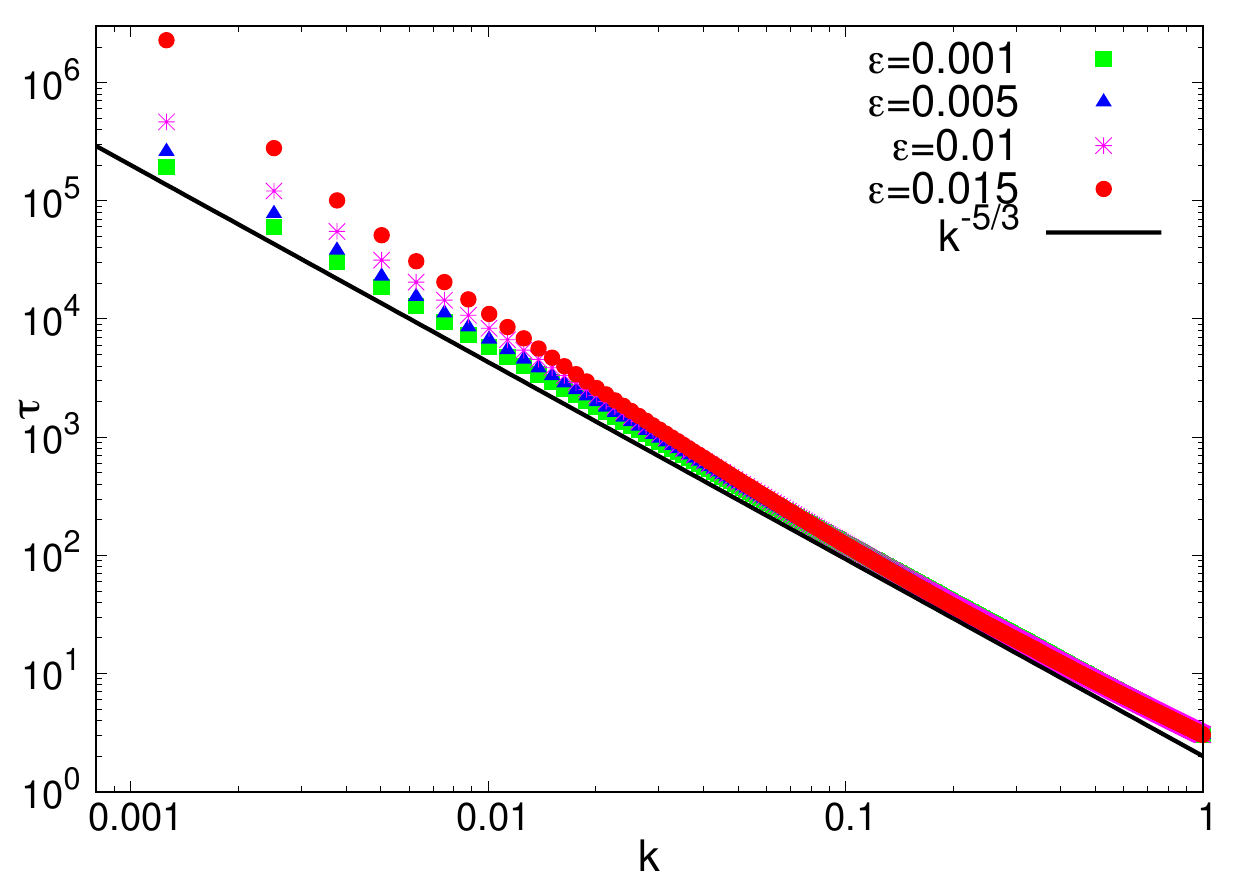}%
\caption{\small Relaxation time, $\tau_k$, as a function of $k$ for $g(k)$ given in (\ref{pert}) with $m=2$, $\mu=0.00001$ and different amplitudes of the perturbation as indicated in the legend. A line, $k^{-5/3}$, resulting from the analytical  prediction for $\epsilon=0$ is also plotted as a reference. }
\label{fig:supp1}
\end{figure}
\begin{figure}
\includegraphics[width=0.5\textwidth]{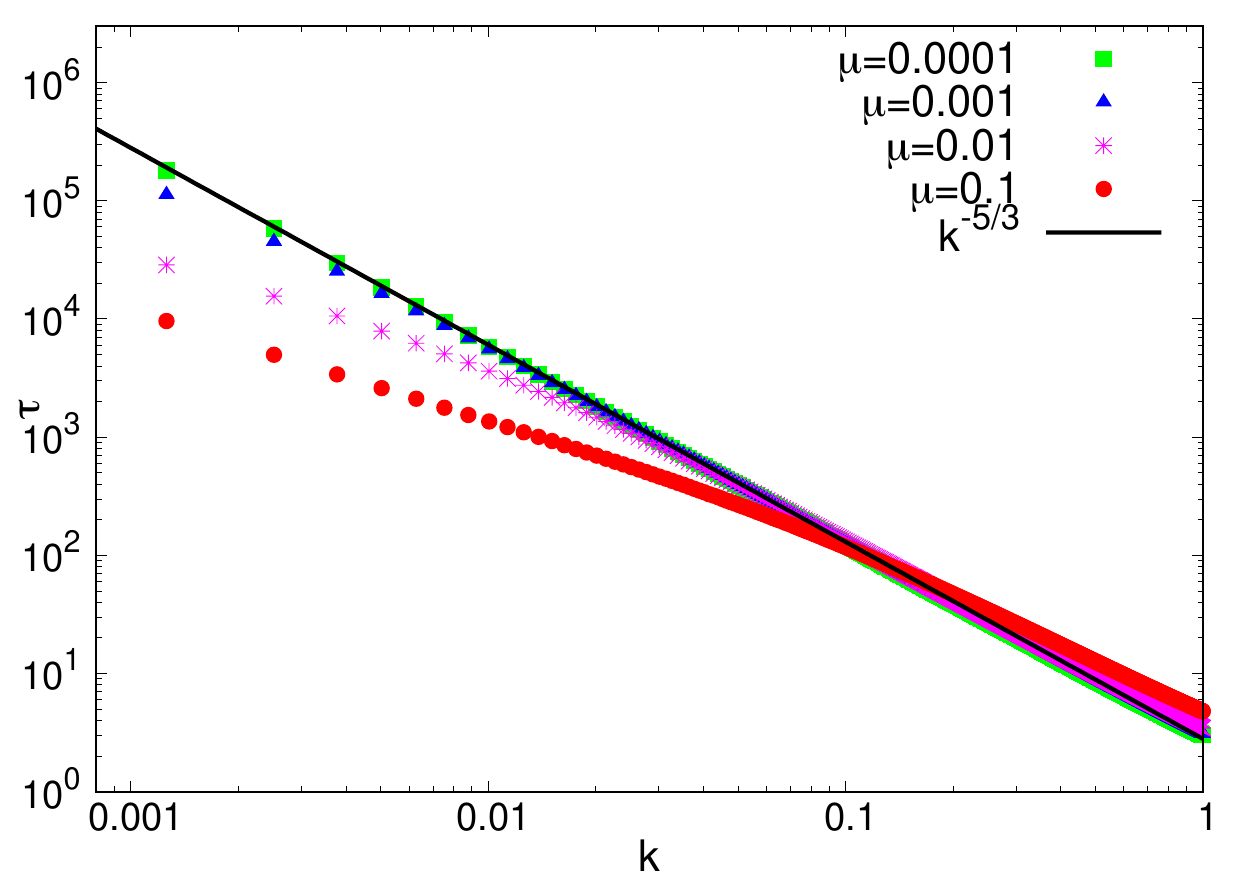}%
\caption{\small Relaxation time, $\tau_k$, as a function of $k$ for $g(k)$ given in (\ref{pert})  with $m=2$, $\epsilon=0.001$ and different values of the chemical potential as indicated in the legend. A line, $k^{-5/3}$, resulting from the analytical  prediction for $\epsilon=0$ is also plotted as a reference.  }
\label{fig:supp2}
\end{figure}
In \cite{pereverzev2003fermi}, the behaviour $\tau_k$ as a function of $k$ has been calculated analytically for $n(k)=n_{RJ}(k)$ with $\mu=0$, in the limit of  small $k$; the result obtained is the following:
\begin{equation}
\tau_k\sim k^{-5/3} \, .
\end{equation}
In order to verify the prediction and test its robustness, we have studied the above integral numerically. We have perturbed the Rayleigh-Jeans distribution in (\ref{RJ}) as follows 
\begin{equation}
n_{RJ}(k)=\frac{T}{\omega(k)+\mu}(1+\epsilon g(k)) \, .
\end{equation}
with
\begin{equation}
g(k)=\cos(m k)
\label{pert}
\end{equation}
and we have computed the integral as a function of $k$ for different values of $\epsilon$, $m$ and $\mu$ ($T$ does not play a role in the determination of the scaling of the integral, as it is just a multiplicative factor).
In Figure \ref{fig:supp1} we show $\tau_k$ for  $m$=2, $\mu=0.00001$ and  $\epsilon$ = 0.001, 0.005, 0.01,0.015. 
In the same plot we also show  the $k^{-5/3}$ line as a reference.
The plot clearly shows that, while the analytical prediction is well reproduced by the numerics in the limit of small $\epsilon$, as the amplitude of the perturbation is slightly increased, departure from the theoretical predictions are observed.  
Another instructive case that we present is related to the role of the chemical potential. In Figure \ref{fig:supp2} we show the relaxation time for $m=2$ and $\epsilon=0.001$ for different values of $\mu$. As the latter parameter increases, then evident departures from the theoretical prediction are observed.

\begin{figure}
\includegraphics[width=0.5\textwidth]{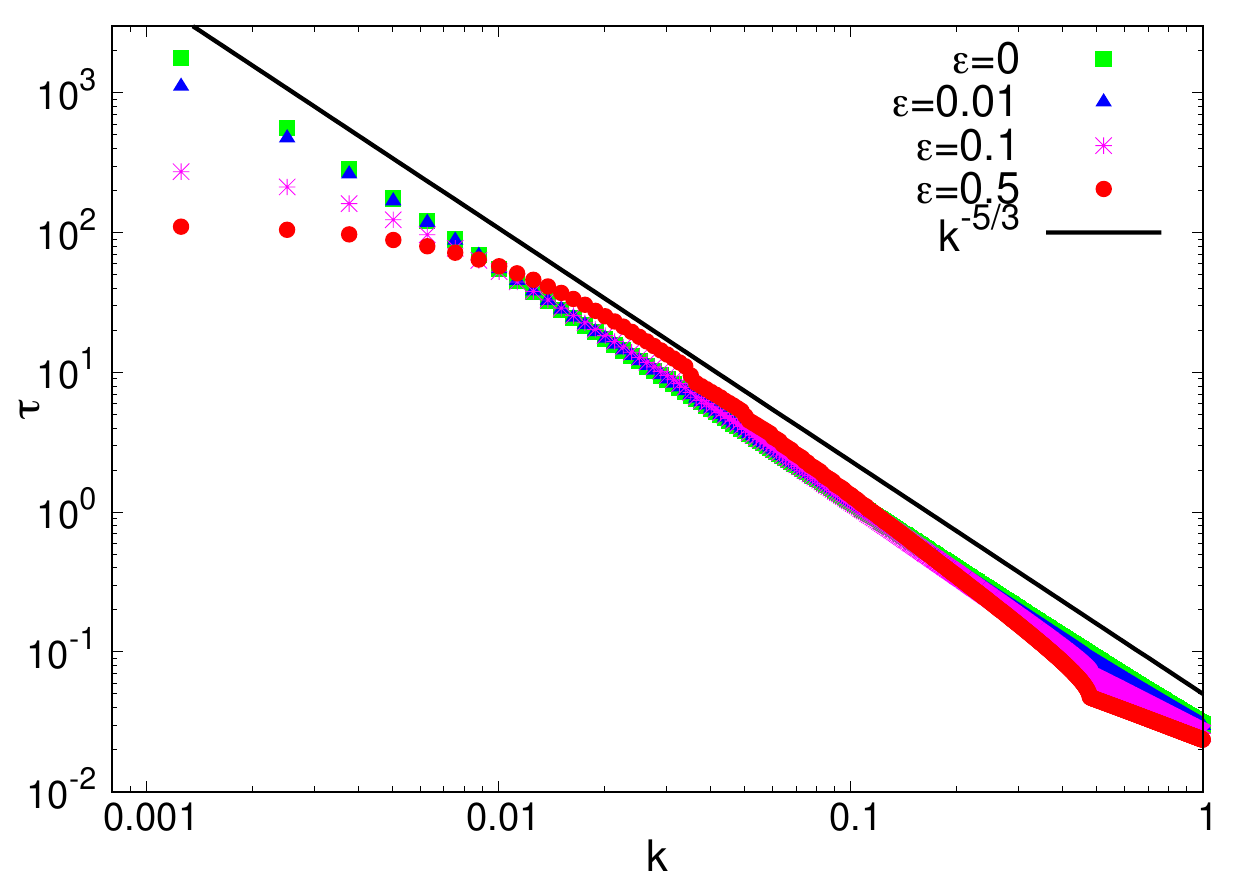}%
\caption{\small Relaxation time, $\tau_k$, as a function of $k$ for $g(k)$ given in (\ref{theta}) with $\mu=0$ and different values of $\epsilon$ as indicated in the legend. A line, $k^{-5/3}$, resulting from the analytical  prediction for $\epsilon=0$ is also plotted as a reference.  }
\label{fig:supp3}
\end{figure}

We have also considered another type of perturbation which corresponds to a condition in which phonons moving towards the thermostat at lower temperature are in equipartition at some temperature, while those moving towards the other thermostat are in equipartition at a lower temperature. To some extents, this is what has been observed in our numerical simulations of the $\beta$-FPUT system. From a mathematical point of view it corresponds to the following perturbation
\begin{equation}
g(k)=\Theta(-x+\pi)-\Theta(x-\pi),
\label{theta}
\end{equation}
with $\Theta(x)$ the Heaviside Step function ($\mu$ was set to 0 in these simulations); in this case  $\epsilon$ measures the relative jump in temperature. The results are shown in \ref{fig:supp3}: as the jump in temperature increases, deviations from the predicted scaling is observed.

In conclusion, the numerical simulations we have presented give evidence that even small deviations from a pure energy equipartition distribution may lead, in the calculation of the relaxation time $ \tau_k $, to substantial deviations from the analytical prediction. Now, it is well known that nonequilibrium conditions quite generally lead low frequency modes to substantially deviate from equipartition \cite{de2006hydrodynamic,conti2013effects,baiesi2020possible}.

\bibliographystyle{apsrev4-1}
%

\end{document}